\documentclass[aps,prl,superscriptaddress,twocolumn]{revtex4}
\usepackage{amsmath,epsfig}
\usepackage{graphicx}

\def\be{\begin{equation}}
\def\ee{\end{equation}}
\def\bea{\begin{eqnarray}}
\def\eea{\end{eqnarray}}

\def\bfr{{\bf r}}

\def\bfB{{\bf B}}
\def\bfE{{\bf E}}

\def\bfS{{\bf S}}

\def\bfd{{\bf d}}

\def\bfrho{\mbox{\boldmath{$\rho$}}}

\def\Gammapl{\Gamma_{\footnotesize\textrm{pl}}}

\def\bfrho{\mbox{\boldmath $\rho$}}

\def\ztrap{z_{\scriptsize\textrm{trap}}}

\begin{document}
\title{Trapping and manipulation of isolated atoms using nanoscale plasmonic structures}

\author{D.E.\ Chang}
\affiliation{Center for the Physics of Information and Institute
for Quantum Information, California Institute of Technology,
Pasadena, CA 91125}

\author{J.D. Thompson}
\affiliation{Department of Physics, Harvard University, Cambridge,
MA 02138}

\author{H. Park}
\affiliation{Department of Physics, Harvard University, Cambridge,
MA 02138} \affiliation{Department of Chemistry and Chemical
Biology, Harvard University, Cambridge, MA 02138}

\author{V. Vuleti\'{c}}
\affiliation{Department of Physics, MIT-Harvard Center for
Ultracold Atoms, and Research Laboratory of Electronics,
Massachusetts Institute of Technology, Cambridge, Massachusetts
02139}

\author{A.S. Zibrov}
\affiliation{Department of Physics, Harvard University, Cambridge,
MA 02138}

\author{P. Zoller}
\affiliation{Institute for Quantum Optics and Quantum Information
of the Austrian Academy of Sciences, A-6020 Innsbruck, Austria}

\author{M.D.\ Lukin}
\affiliation{Department of Physics, Harvard University, Cambridge,
MA 02138}

\date{\today}

\begin{abstract}
We propose and analyze a scheme to interface individual neutral
atoms with nanoscale solid-state systems.  The interface is
enabled by optically trapping the atom via the strong near-field
generated by a sharp metallic nanotip.  We show that under
realistic conditions, a neutral atom can be trapped with position
uncertainties of just a few nanometers, and within tens of
nanometers of other surfaces. Simultaneously, the guided surface
plasmon modes of the nanotip allow the atom to be optically
manipulated, or for fluorescence photons to be collected, with
very high efficiency. Finally, we analyze the surface forces and
heating and decoherence rates acting on the trapped atom.
\end{abstract}

\maketitle

Much interest has recently been directed towards hybrid systems
that integrate isolated atomic systems with solid-state
devices~\cite{monroe08,andre06,folman02,treutlein06}. These
efforts are aimed at combining the best of both worlds, namely the
excellent coherence and control associated with isolated atoms,
ions and molecules, with the miniaturization and integrability
associated with solid-state devices. A key ingredient for such
integrated devices is the ability to trap, coherently manipulate,
and measure individual cold atoms at distances below
${\sim}100$~nm from solid-state surfaces.

In this Letter, we describe a technique that allows a single atom
to be optically trapped within a nanoscale region above the
surface of a sharp, conducting nanotip.  Under illumination with a
single blue-detuned laser beam, the nanotip behaves as a
``lightning rod'' that generates very large field gradients and an
intensity minimum that can be used to tightly trap an atom.
Simultaneously, the nanotip supports a set of tightly guided
surface plasmon modes to which the trapped atom can very
efficiently couple. Under realistic conditions, the strong
coupling regime can be reached, where the emission rate into the
guided surface plasmons of the nanotip far exceeds that into all
other channels. It has been shown that this regime enables
efficient fluorescence collection and optical manipulation at a
single-photon level~\cite{chang07a,chang07b,akimov07}. Finally, we
analyze in detail surface effects and photon scattering and their
effects on trap lifetimes and atomic coherence times.

\begin{figure*}[t]
\begin{center}
\includegraphics[width=17cm]{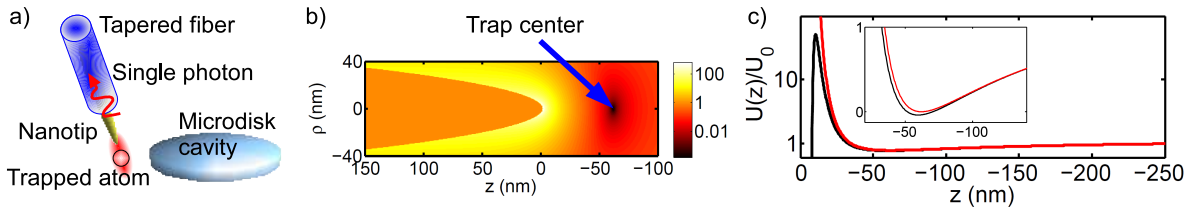}
\end{center}
\caption{a) Schematic of a single atom tightly trapped near a
conducting nanotip. The atom is strongly coupled to single surface
plasmons guided by the nanotip, which can be efficiently converted
to a single photon in a coupled optical fiber, allowing for
efficient manipulation and read-out. The atom can be brought
within tens of nanometers of other surfaces, which allows it to be
interfaced, \textit{e.g.}, with an optical microdisk cavity as
shown here. b) Illustration of a nanotip with $z_{0}=2$~nm and
normalized total field intensity
$|E_{\scriptsize\textrm{total}}/E_0|^2$. c) Typical optical
potential~(red) and total potential~(including van der Waals
potential, black) for a Rb atom trapped near a $z_{0}=2$~nm
nanotip. The potentials are normalized by
$U_0=\hbar\Omega_0^2/\delta$, the optical potential at infinity.
The inset shows the potentials around the trap center in greater
detail.\label{fig:PRLtip}}
\end{figure*}

The trapping technique described in this Letter might enable the
realization of several unique applications. For example, the
nanotrap can be used for deterministic positioning of single atoms
near micro-photonic and nano-photonic structures, such as
micro-toroidal resonators~\cite{vahala03,aoki06} and photonic
crystal cavities~\cite{vuckovic03}~(see Fig.~\ref{fig:PRLtip}a).
Alternatively, the trap can be used for realization of hybrid
quantum systems consisting of single atoms or molecules in the
immediate proximity of charged or magnetized solid-state quantum
systems, to enable direct strong coupling~\cite{treutlein07}.
Finally, a trapped atom might be used as a novel scanning probe
for sensing magnetic or electric fields with nanoscale resolution.
We note that forces associated with metallic systems are being
explored, in the context of optical tweezers for dielectric
objects on surfaces~\cite{righini07} and electro-optical atomic
trapping using nanotubes~\cite{murphy09}. In contrast to the
latter work, our scheme offers an all-optical trapping method, an
open geometry~\cite{maiwald08}, and an efficient mechanism for
optical readout and manipulation.

We first derive the optical trapping potential experienced by an
atom in the vicinity of the nanotip, whose surface is
parameterized by a paraboloid of revolution with rotational axis
along $z$, $z(\rho)=-z_{0}+\rho^2/4z_0$. Here $z_0$ characterizes
the curvature of the tip and $\rho=\sqrt{x^2+y^2}$ is the radial
coordinate~(see Fig.~\ref{fig:PRLtip}b for an illustration of a
tip with $z_0=2$~nm).  The end of the nanotip is thus located at
$z=-z_0$~(the offset from $z=0$ is conventional in the
transformation to paraboloidal coordinates that facilitates
analysis). We consider the total field produced by a plane wave
incident upon the nanotip from the far field,
$\bfE_{\scriptsize\textrm{inc}}(\bfr)=E_{0}e^{ik_{L}x-i{\omega}_{L}t}\hat{z}$,
which is polarized along the nanotip axis. While an exact
analytical solution cannot be obtained for this geometry, the near
field around a sub-wavelength nanotip can be approximated using
electrostatic equations that do admit analytical
solutions~\cite{chang07a}. Within this approximation, the total
field in the region outside the nanotip is given by
\be
\bfE_{\scriptsize\textrm{total}}=E_{0}\left(1+\frac{z_0}{r}\left(\epsilon_{L}-1\right)\right)\hat{z}+\frac{E_{0}z_0}{r(r-z)}\left(1-\epsilon_L\right)\bfrho,\label{eq:Ertrap}
\ee
while the field inside the nanotip is uniformly
$\bfE_{\scriptsize\textrm{total}}=E_{0}\hat{z}$. Here
$\epsilon_{L}\equiv\epsilon(\omega_L)$ is the dimensionless
electric permittivity of the nanotip at the laser frequency~(we
assume that the surrounding material is vacuum, $\epsilon=1$) and
$r=\sqrt{\rho^2+z^2}$. When the nanotip is conducting and far
below its plasma resonance, such that $\epsilon_{L}{\ll}-1$, the
field $\bfE_{\scriptsize\textrm{total}}=\epsilon_{L}E_{0}\hat{z}$
at the tip end is greatly enhanced and out of phase relative to
the incident field. This is essentially the ``lightning rod''
effect of a good conductor~\cite{jackson99}. Far away from the tip
the field must relax back to $E_0$ and the total field passes zero
at ${\ztrap}=z_{0}(\epsilon_{L}-1)$. A small residual field
remains if $\epsilon_{L}$ has a small imaginary component
($\epsilon_L{\approx}-30+0.4i$ in silver at wavelength
$\lambda_L=780$~nm~\cite{johnson72}). The induced field also
varies rapidly near the tip, on a length scale characterized by
$|{\ztrap}|$ that can be much smaller than the optical wavelength.

For a simple two-level atom, the field minimum provides a trapping
potential when the laser is blue-detuned from the transition
frequency $\omega_a$~($\delta\equiv\omega_L-\omega_a>0$), such
that the atomic polarizability is negative. Expanding the fields
linearly around the trap center, the potential corresponds to that
of a harmonic oscillator, whose trapping frequency $\omega_{T,z}$
along $z$ is given by
\be \hbar\omega_{T,z}=
2\sqrt{\frac{\hbar\Omega_0^2}{\delta}E_{R}'}, \ee
where $E_{R}'=E_{R}(k_{a}{\ztrap})^{-2}$ is an effective
``enhanced'' recoil energy relative to the recoil energy
$E_R=\hbar^{2}k_a^2/2m$ in free space, $m$ is the mass of the
atom, $k_a=\omega_a/c$, and $\Omega_0$ is the Rabi frequency
associated with the incident field amplitude. The ground-state
uncertainty of the trap along $z$ is
$a_{z}=\sqrt{\hbar/2m\omega_{T,z}}$, while the trap frequency in
the radial directions is $\omega_{T,\rho}=\omega_{T,z}/2$. Note
that the field gradients created by the nanotip strongly enhance
$E_{R}'$, such that larger trap frequencies
$\omega_{T,z}{\propto}1/|{\ztrap}|$ can be obtained for a given
input intensity.

\begin{figure}[t]
\begin{center}
\includegraphics[width=8cm]{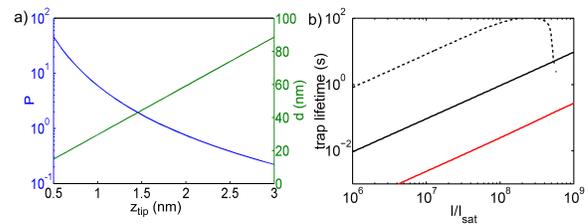}
\end{center}
\caption{a) Purcell factor $P$ (averaged over dipole orientations,
blue curve) and trap distance to surface $d$~(green curve) for a
${}^{87}$Rb atom trapped near a silver nanotip, as a function of
$z_0$, in absence of van der Waals forces. b) Trap lifetime for
various values of trapping frequency $\omega_{T,z}$ and tip
curvature $z_0$. The incident trap laser intensity is plotted
along the horizontal axis, while the detuning is varied to
maintain a given value of $\omega_{T,z}$. The black dashed~(solid)
curve corresponds to $z_0=3$~nm and trap frequency of
$\omega_{T,z}=10$~$(100)$~MHz, while the red curve corresponds to
$z_{0}=1$~nm and
$\omega_{T,z}=100$~MHz.\label{fig:trapproperties}}
\end{figure}

An atom trapped in this nanoscale region can be optically
manipulated and read out with near unit efficiency via efficient
coupling of the atom to guided surface plasmons~(SPs) that
propagate along the nanotip surface. Following the ideas of
Ref.~\cite{chang06}, a large coupling strength between a single
SP~(\textit{i.e.}, a single photon) and single atom results when
the atom is placed within the SP evanescent field, due to the
sub-diffraction limit confinement of the SPs. This effect yields
an enhanced spontaneous emission rate $\Gammapl$ into the SPs over
the rate $\Gamma'$ into all other channels, which can be
characterized by an ``effective Purcell factor''
$P=\Gammapl/\Gamma'$. To illustrate this effect, the Purcell
factor for an atom~(emission wavelength at $\lambda_a=780$~nm)
trapped at position $z={\ztrap}$ near a silver nanotip is plotted
in Fig.~\ref{fig:trapproperties}a as a function of $z_0$. It can
be seen that the strong coupling regime $P>1$ can be achieved over
a realistic range of $z_0$. Furthermore, the strong coupling is
broadband and associated primarily with the small tip size, and
thus no special tuning of the nanotip is required to achieve the
strong coupling for some particular atom.

For realistic parameters, the distance $d=|\epsilon_{L}|z_0$
between the trap center and tip surface is on the order of tens of
nanometers~(see Fig.~\ref{fig:trapproperties}a), and thus surface
effects can play an important role in the trap characteristics.
Here we identify and analyze several prominent surface effects --
an attractive van der Waals force from the nanotip and ``patch
potentials'' caused by adatoms that modify the total potential
experienced by the atom, and magnetic field fluctuations caused by
``polarization noise'' in the nanotip that induce both motional
heating and hyperfine state flips in a multilevel atom. The van
der Waals force can be calculated classically based on the
interaction between an oscillating dipole $\bfd$ at some point
$\bfr$ and its own reflected field~\cite{chance75}. This
calculation is valid when $k_{a}d{\lesssim}1$~(such that
retardation is not important), which is true for our cases of
interest. Taking the known result for the field reflected from a
nanotip~\cite{chang07a}, the van der Waals potential as
$z{\rightarrow}-z_0$ is given for a two-level atom by
$U_{\scriptsize\textrm{vdW}}(z){\approx}-\frac{3\hbar\Gamma_0}{32k_{a}^{3}d^3}$,
where $\Gamma_0$ is the free-space spontaneous emission rate.
$U_{\scriptsize\textrm{vdW}}$ is attractive and diverges with the
usual $d^{-3}$ scaling as the distance to the surface $d$
approaches zero. For sufficiently weak optical potentials, the
total potential
$U_{\scriptsize\textrm{opt}}+U_{\scriptsize\textrm{vdW}}$ may
cease to support a trapping minimum away from the surface. The
condition for a trap to exist is approximately
\be
\frac{\Omega_0^2}{\delta}{\gtrsim}\frac{9\Gamma_0}{32(k_{a}|{\ztrap}|)^3},\label{eq:trapexistence}
\ee
\textit{i.e.}, the strength of the laser potential should roughly
exceed that of the van der Waals force at the trap position. Even
if the condition above is satisfied, some probability remains for
the atom to tunnel from the local trapping minimum to the surface.
However, the tunneling rate is exponentially suppressed with
barrier height and can be ignored once
Eq.~(\ref{eq:trapexistence}) is even moderately satisfied.

A second correction to the total potential occurs if additional
adatoms~(in addition to the trapped atom) become adsorbed to the
nanotip surface~\cite{mcguirk04}. Each adatom forms a static
dipole moment $p_0$ due to its electronic wave function being
pulled into or away from the surface, thus producing a small
static electric field. The total static field $E_p$ in turn
creates an additional ``patch'' potential for the trapped atom,
$U_{p}=-(1/2){\alpha_s}E_p^2$, where $\alpha_s$ is the atomic
static polarizability.  Assuming that a uniform monolayer of
adatoms substitute themselves over some portion of the nanotip
surface, it is straightforward to show that the maximum
force~(\textit{i.e.}, in a worst-case scenario) at a distance $d$
away from the nanotip is given by $F_{p}(d)\lesssim
0.1\frac{p_0^{2}z_0^2\alpha}{\epsilon_0^{2}d^{5}a^4}$, where $a$
is the lattice spacing of the nanotip material.  For typical
dipole moments of $p_{0}{\sim}1$~Debye, this force shifts the trap
center only by an amount comparable to the size of the
ground-state wavepacket near the minimum intensities needed to
overcome the van der Waals
potential~(Eq.~(\ref{eq:trapexistence})). Thus from this point
forward we will ignore the possibility of patch potentials in our
calculations.

Eq.~(\ref{eq:trapexistence}) predicts that the minimum amount of
incident laser intensity needed to support a trap rises rapidly
with decreasing nanotip size. On the other hand, for sufficiently
large intensities the laser power absorbed by the nanotip, as
determined by the imaginary part of $\epsilon_L$, will cause it to
melt. Assuming that the nanotip has a good thermal contact
conductance~(\textit{e.g.}, comparable to achievable values for
wires in atom chips~\cite{groth04}) with some substrate, we
estimate that incident laser intensities exceeding 10
mW/$\mu$m${}^2$ can be used for silver nanotips at
$\lambda_L=780$~nm~\cite{heatingfootnote1}. These two
considerations set a lower bound for the tip size $z_{0}$ of
around hundreds of picometers. An upper bound for $z_{0}$ is set
by the validity of our electrostatic calculations. Specifically,
in a subwavelength region around the end of the tip, the tip
profile must appear ``sharp''~(as defined by having a large aspect
ratio of $z/2\rho$), and the trap distance should satisfy
$k_{a}d{\lesssim}1$, which places an upper limit to $z_0$ of
several nanometers.

We now discuss the limitations on atomic coherence times and trap
lifetimes. First, the proximity of the trap to the surface makes
the atom susceptible to magnetic field noise $\bfB_{N}$ induced by
material losses in the nanotip. This field noise couples to the
electron spin via the Hamiltonian
$V=-\mu_{B}g_{S}\bfS\cdot\bfB_{N}(\bfr,t)$, resulting in
incoherent transitions between ground-state hyperfine levels and
jumps between trap motional states. Here $\mu_B$ is the Bohr
magneton and $g_S$ is the electron spin g-factor. An analytical
solution for $\bfB_N$ cannot be found for the nanotip. However, to
estimate its effect, we can consider an atom sitting a distance
$d$ above a semi-infinite substrate of the same permittivity as
the nanotip. The hyperfine transition
rate~$\Gamma_{\Delta{F},\scriptsize\textrm{mag}}$ and motional
jump rate $\Gamma_{\scriptsize\textrm{jump,mag}}$ due to magnetic
noise in this case are $\Gamma_{\Delta{F},\scriptsize\textrm{mag}}
\propto \frac{(\mu_{0}\mu_{B}g_{S})^2}{\hbar^{2}\rho{d}}k_{B}T,
\Gamma_{\scriptsize\textrm{jump,mag}}\propto\Gamma_{\Delta{F},m}(a_z/d)^2$~\cite{henkel99}
where $\rho$ is the resistivity of the nanotip. We note that the
semi-infinite substrate over-estimates the amount of polarizable
material and that for realistic tips the noise should be reduced
by a factor of order ${\sim}(z_{0}/z_{trap})^2$. The hyperfine
transitions result only in a loss of internal atomic coherence,
since all hyperfine states can be trapped in the optical fields.
In the following we assume that the nanotip roughly sits at room
temperature, $T{\sim}300$~K.

Analogous processes occur due to inelastic scattering of photons
from the trapping field. Because of the tight trap confinement,
the change in motional state primarily consists of events where a
single phonon is added or subtracted, in analogy with heating of
ions in the Lamb-Dicke limit~\cite{leibfried03}. The transition
rates can generally be obtained by second-order perturbation
theory~\cite{wineland79,cline94}. For a two-level system, we find
a jump rate
\be
\Gamma_{\scriptsize\textrm{jump,opt}}\approx\Gamma_{\scriptsize\textrm{total}}^{(z)}\frac{E_R'}{\hbar\omega_{T,z}}\frac{\Omega_0^2}{\delta^2},
\ee
where $\Gamma_{\scriptsize\textrm{total}}^{(z)}$ denotes the total
spontaneous emission rate for a dipole oriented along the nanotip
axis. Note that the enhanced recoil energy $E_R'$ yields a larger
heating rate as compared to free space. Photon scattering also
results in hyperfine transitions, which we calculate using
analogous techniques~\cite{cline94}. An additional source of
heating is laser shot noise, which causes fluctuations in the trap
frequency $\omega_{T}$. For a laser beam focused to
${\sim}\lambda^2$, however, this heating is smaller than
$\Gamma_{\scriptsize\textrm{jump,opt}}$ by a factor
${\sim}(a_z/d)^2$.

As a numerical example, we now consider the trapping of individual
${}^{87}$Rb atoms~($\lambda_a{\sim}780$~nm for the D2 line,
$\Gamma_0{\sim}38$~MHz, saturation intensity
$I_{\scriptsize\textrm{sat}}{\sim}1.7$~mW/cm${}^2$) near a silver
nanotip. For the nanotip heating rates calculated previously,
traps with laser intensities of up to
$I{\sim}10^{9}I_{\scriptsize\textrm{sat}}$ can be realized. In
these examples, both the complex value of $\epsilon_L$ and the
multilevel atomic structure of Rb have been fully accounted for.
For the latter consideration, the optical interactions include the
atomic fine structure, and are averaged assuming that all magnetic
states $m_{F}$ are trapped with equal populations. In
Fig.~\ref{fig:trapproperties}b we plot the trap lifetime for
various values of $\omega_{T,z}$ and $z_{0}$. The incident field
intensity is shown along the horizontal axis, while the detunings
are varied to yield the desired values of $\omega_{T,z}$. The
black dashed~(solid) curve corresponds to a nanotip of $z_0=3$~nm
and trap frequency of $\omega_{T,z}=10$~(100)~MHz, while the red
curve corresponds to $z_0=1$~nm and $\omega_{T,z}=100$~MHz. Note
that $\omega_{T,z}=100$~MHz corresponds to a ground-state
localization of $a_{z}{\sim}2$~nm. The van der Waals force has
been accounted for in calculating the total potential, and the
trap lifetime is the time it takes for the energy of the atom to
exceed the depth of the total potential. For the
$\omega_{T,z}=10$~MHz curve, trap lifetimes exceeding ${\sim}1$~s
can be readily achieved. At the same time, spin flip
times~(dominated by magnetic field noise at large detunings
$\delta{\sim}10^{6}\Gamma_0$) are conservatively calculated to be
around $10$~ms using the results obtained for a semi-infinite
substrate; however, by estimating a correction based on the small
solid angle actually spanned by the nanotip, we believe that times
on the order of a second are possible. In the regime where the van
der Waals potential does not perturb the trap significantly, the
trapped atom sits 90~(30)~nm from the tip surface for a tip
curvature of $z_{0}=3$~(1)~nm, giving a corresponding Purcell
factor of $P{\sim}0.2$~(6) when averaged over dipole orientations.

Loading the nanotip trap can be accomplished starting with an atom
initially trapped within a few-micron vicinity of the nanotip in a
separate, far-field red-detuned optical dipole trap. Suppose in
particular that the far-field trapping beam is polarized
perpendicular to the nanotip axis~(say along $\hat{x}$). A similar
analysis as above shows that the total field for a plane wave
along the nanotip axis is given by
$\bfE_{\scriptsize\textrm{total}}=E_{0}\left(1+\frac{1-\epsilon_L}{1+\epsilon_L}\frac{z_0}{|z|}\right)\hat{x}$.
Thus, for this polarization, the incident field is only modified
at very close distances to the tip of order
$z_{\perp}{\sim}z_{0}(\epsilon_{L}-1)/(\epsilon_L+1){\ll}|{\ztrap}|$,
\textit{i.e.}, the nanotip has minimal effect on the far-field
trap. The atom can then be loaded into our system by adiabatically
switching on~(off) the nanotip~(far-field) trap. The different
responses of the two polarizations ensures that the trapping
potentials of the red- and blue-detuned beams do not simply cancel
each other out in the loading process.

In summary, we have described a technique that allows for the
nanoscale trapping and efficient optical manipulation of single
atoms on a chip. Such a trap is expected to display long trap
lifetimes and atomic coherence times, and its open geometry and
large depth allow the trapped atom to be brought into close
proximity~(${\sim}50$~nm) of other surfaces as well. The
combination of these features potentially opens up many exciting
opportunities. For example, the nanotip could be used to trap
atoms within the evanescent field modes of optical resonators such
as whispering-gallery mode resonators~\cite{vahala03,aoki06} and
photonic crystal cavities~\cite{vuckovic03}. The nanotip may also
be used as a scanning tip for weak-field sensing near surfaces. In
magnetometry, for example, the field sensitivity will be
determined by the atomic spin coherence time $T_2$. Estimating
coherence times of $T_{2}{\sim}1$~s yield ultimate magnetic field
sensitivities of
$\delta{B}{\sim}\hbar/g_{S}\mu_{B}\sqrt{T_2}{\sim}20$~pT$/\sqrt{\textrm{Hz}}$,
which compare favorably with spins localized in solid
state~\cite{taylor08}. Furthermore, the tight trapping is expected
to give rise to novel atomic interactions. If two atoms are
trapped simultaneously, the ground-state uncertainties can be made
comparable to the length scale over which they experience a van
der Waals interaction~\cite{bolda02}. In this regime, optical
forces could directly ``push'' on a molecule and alter its
properties and dynamics. Also, arrays of nanotips could form
optical lattices with very small lattice constants, enabling the
exploration of novel many-body physics~\cite{bloch08}. Finally,
these ideas could also be extended to other systems of interest,
such as polar molecules~\cite{andre06} or ions~\cite{leibfried03}.

This work was supported by the NSF, Harvard-MIT CUA, DARPA, and
Packard Foundation. DEC acknowledges support from the Gordon and
Betty Moore Foundation through Caltech's Center for the Physics of
Information, and the National Science Foundation under Grant No.
PHY-0803371.
\bibliographystyle{apsrev}
\bibliography{../bibsalpha}

\begin{thebibliography}{28}
\bibitem[{\citenamefont{{This calculation of the maximum intensity is
  consistent with preliminary experiments involving silver nanowires, where
  green laser light with intensities exceeding $10$~mW/$\mu$m${}^2$ have been
  found to not melt the nanowires.}}()}]{heatingfootnote1}
\bibinfo{author}{\bibnamefont{{This calculation of the maximum intensity is
  consistent with preliminary experiments involving silver nanowires, where
  green laser light with intensities exceeding $10$~mW/$\mu$m${}^2$ have been
  found to not melt the nanowires.}}}
\end{thebibliography}

\end{document}